# Distribution of Maximum Earthquake Magnitudes in Future Time Intervals, Application to the Seismicity of Japan (1923-2007)

V.F. Pisarenko[1], D. Sornette[2-4] and M.V. Rodkin[5].

[1]International Institute of Earthquake Prediction Theory
and Mathematical Geophysics
Russian Ac. Sci., Profsoyuznaya 84/32, Moscow 117997, Russia
[2]ETH Zurich, D-MTEC, Kreuzplatz 5
CH-8032 Zurich, Switzerland
[3]ETH Zurich, Department of Earth Sciences
[4]Institute of Geophysics and Planetary Physics
University of California, Los Angeles, California 90095
[5]Geophysical Center of Russian Academy of Sciences, Molodejnaya 3, Moscow, Russia

(15.08.2009)

*Abstract:* We modify the new method for the statistical estimation of the tail distribution of earthquake seismic moments introduced by Pisarenko et al. [2009] and apply it to the earthquake catalog of Japan (1923-2007). The method is based on the two main limit theorems of the theory of extreme values and on the derived duality between the Generalized Pareto Distribution (GPD) and Generalized Extreme Value distribution (GEV). We obtain the distribution of maximum earthquake magnitudes in future time intervals of arbitrary duration $\tau$. This distribution can be characterized by its quantile $Q_q(\tau)$ at any desirable statistical level $q$. The quantile $Q_q(\tau)$ provides a much more stable and robust characteristic than the traditional absolute maximum magnitude $M_{max}$ ($M_{max}$ can be obtained as the limit of $Q_q(\tau)$ as $q \to 1$, $\tau \to \infty$). The best estimates of the parameters governing the distribution of $Q_q(\tau)$ for Japan (1923-2007) are the following:
$\xi_{GEV} = -0.1901 \pm 0.0717$; $\mu_{GEV}(200) = 6.3387 \pm 0.0380$; $\sigma_{GEV}(200) = 0.5995 \pm 0.0223$; $Q_{0.90,GEV}(10) = 8.34 \pm 0.32$. We also estimate $Q_q(\tau)$ for a set of $q$-values and future time periods in the range $1 \leq \tau \leq 50$ years from 2007. For comparison, the absolute maximum estimate $M_{max,GEV} = 9.57 \pm 0.86$ has a scatter more than twice that of the 90% quantile $Q_{0.90,GEV}(10)$ of the maximum magnitude over the next 10 years counted from 2007.

*1-Introduction*

The present work has two goals: (i) adapt the method suggested in [Pisarenko et al. 2009] for the statistical estimation of the tail of the distribution of earthquake magnitudes to catalogs in which earthquake magnitudes are reported in discrete values, and (ii) apply the obtained method to the JMA magnitude catalog of Japan (1923-2007) to estimate the maximum possible magnitude and other measures characterizing the tail of the distribution of magnitudes.

The method of [Pisarenko et al. 2009] is a continuation and improvement of the technique suggested in [Pisarenko et al. 2008]. Both rely on the assumption that the distribution of earthquake magnitudes is limited to some maximum value $M_{max}$, which is itself probably significantly less than the absolute limit imposed by the finiteness of the Earth. This maximum value $M_{max}$ may reflect the largest possible set of seismo-tectonic structures in a given tectonic region that can support an earthquake, combined with an extremal occurrence of dynamical energy release per unit area. The simplest model embodying the idea of a maximum magnitude is the truncated Gutenberg-Richter (GR) magnitude distribution truncated at $M_{max}$:

$$F(m) = C[1 - 10^{-b(m-m_0)}]; \qquad m_0 \leq m \leq M_{max}, \qquad (1)$$

where $F(m)$ is the cumulative probability distribution of earthquake magnitudes, $b$ is the slope parameter, $m_0$ is the lower known threshold above which magnitudes can be considered to be reliably recorded, $M_{max}$ is the maximum possible magnitude, and C is the normalizing constant (which depends on the unknown parameters $b$ and $M_{max}$) [Cosentino et al., Kijko and Sellevol, 1989; 1992; Pisarenko et al., 1996; Kijko, 2004]. The parameter $M_{max}$ is a priori very convenient for building engineering and for insurance business. However, the multiple attempts to use $M_{max}$ has definitely shown that this parameter is unstable with respect to minor variations of the catalogs, and in particular for incomplete regional catalogs, a rather common situation in seismology. The parameter $M_{max}$ is thus an unreliable measure of the largest seismogical risks. The truncated GR model (1) can be contrasted with the various modifications of the Gutenberg-Richer law stretching to the infinity, which impose a finite-size constraint only on the statistical average of the energy released by earthquakes (see e.g . [Sornette et al., 1996; Kagan, 1999; Kagan and Schoenberg, 2001], but they contradict to finiteness of seismogenic structures in the Earth and did not get a universal acceptance.

The chief innovation, introduced in [Pisarenko et al. 2009] and that we extend here, is to combine the two main limit theorems of Extreme Value Theory (*EVT*) that allow us to derive the distribution of *T*-maxima (maximum magnitude occurring in sequential time intervals of duration $T$) for arbitrary $T$. This distribution enables one to derive any desired statistical characteristic of the future $T$-maximum. The two limit theorems of EVT correspond respectively to the Generalized Extreme Value distribution (*GEV*) and to the Generalized Pareto Distribution (*GPD*). Pisarenko et al. [2009] have established the direct relations between the parameters of these two distributions. The duality between the *GEV* and *GPD* provides a new way to check the consistency of the estimation of the tail characteristics of the distribution of earthquake magnitudes for earthquake occurring over arbitrary time interval.

Instead of focusing on the unstable parameter $M_{max}$ we suggest a new, stable and convenient characteristic $M_{max}(\tau)$ – defined as the maximum earthquake that can be recorded over a future time interval of duration $\tau$. The random value $M_{max}(\tau)$ can be described by its distribution function, or by its quantiles $Q_q(\tau)$, that are, in contrast with $M_{max}$, stable and robust characteristic. Besides, if $\tau \to \infty$, then $M_{max}(\tau) \to M_{max}$ with probability one. The methods of calculation of $Q_q(\tau)$ are exposed below. In particular, we can estimate $Q_q(\tau)$ for, say, $q = 10, 5$ and *1%*, as well as for the median *(q = 50%)* for any desirable time interval $\tau$. These methods are illustrated below on the magnitude catalog of the Japanese Meteorological Agency (JMA), over the time period 1923-2007, for magnitudes $m \geq 4.1$.

*2-The method*

According to the theory of extreme values, the limit distribution of maxima can be obtained in two ways. The first one, sometimes called the "peak over threshold" method, consists in increasing a threshold $h$ above which observations are kept. Then, the distribution of event sizes, which exceed $h$, tends (as $h$ tends to infinity) to the Generalized Pareto Distribution (GPD). The GDP depends on two unknown parameters *(ξ, s)* and on the known threshold $h$ (see e.g. [Embrechts et al., 1997]). For the case of random values where are limited from above, the GPD can be written as follows:

$$GPD_h(x | \xi, s) = 1 - [1 + (\xi/s) \cdot (x - h)]^{-1/\xi}, \qquad \xi < 0; s > 0; \quad h \leq x \leq h - s/\xi. \qquad (2)$$

Here, $\xi$ is the form parameter, $s$ is the scale parameter and the combination $h - s/\xi$ represents the uppermost magnitude, that we shall denote $M_{max}$:

$$M_{max} = h - s/\xi, \quad \xi < 0. \qquad (3)$$

We shall consider only this case of a finite $M_{max}$, to capture the finiteness of seismo-tectonic structures in the Earth, as discussed in the introduction.

The second way consists in selecting directly the maxima occurring in sequences of *n* successive observations $M_n = max(m_1, ..., m_n)$, and in studying their distribution as *n* goes to infinity. In accordance with the main theorem of the theory of extreme values (see e.g. [Embrechts et al., 1997]), this distribution, named the Generalized Extreme Value distribution (GEV), can be written (for the case of random values limited from above) in the form:

$$GEV(x \mid \zeta, \sigma, \mu) = exp(-[1 + (\zeta/\sigma) \cdot (x - \mu)]^{-1/\zeta}, \qquad \zeta < 0; \sigma > 0; \quad x \leq \mu - \sigma/\zeta. \qquad (4)$$

The conditions guaranteeing the validity of these two limit theorems include the regularity of the original distributions of magnitudes in their tail and boil down to the existence of a *non-degenerate* limit distribution of $M_n$ after a proper centering and normalization.

We shall study the maximum magnitudes occurring in time interval *(0, T)*. We assume that the flow of main shocks is a Poissonian stationary process with some intensity $\lambda$. This property for main shocks was studied and confirmed in Appendix A of [Pisarenko et al., 2008], for the Harvard catalog of seismic moments over the time period 01.01.77 – 20.12.04. The term «main shock» refers here to the events remaining after using a suitable declustering algorithm (see [Pisarenko et al., 2008; 2009] and below). Given the intensity $\lambda$ and the duration *T* of the time window, the average number of observations (main shocks) within the interval *(0, T)* is equal to $<n> = \lambda T$. For $T \rightarrow \infty$, the number of observations in *(0, T)* tends to infinity with probability one and we can use (4) as the limit distribution of the maximum magnitudes $m_T$ of the main shocks occurring in time interval *(0, T)* of growing sizes [Pisarenko et al., 2008].

Pisarenko et al. [2009] have shown that, for a Poissonian flow of main shocks, the two limit distributions, the GPD given by (2) and the GEV given by (4), are related in a simple way. We briefly summarize the main points, and refer to [Pisarenko et al., 2009] for details. If the random variable (rv) *X* has the GPD-distribution (2) and one takes the maximum of a random sequence of observations $X_k$,

$$M_T = max(X_1, ..., X_\nu), \qquad (5)$$

where $\nu$ is a random Poissonian value with parameter $\lambda T$, with $\lambda T >> 1$, then $M_T$ has the GEV-distribution (4) with the following parameters:

$$\zeta(T) = \xi ; \qquad (6)$$

$$\sigma(T) = s \cdot (\lambda T)^{\xi} ; \qquad (7)$$

$$\mu(T) = h - (s/\xi) \cdot [1 - (\lambda T)^{\xi}] . \qquad (8)$$

These expressions are valid up to small terms of order $exp(-\lambda T)$, which are neglected.

The inverse is true as well: if $M_T = max(X_1, ..., X_\nu)$ has the GEV-distribution (4) with parameters $\zeta, \sigma, \mu$, then the original distribution of $X_k$ has the GPD-distribution (2) with parameters:

$$\xi = \zeta; \qquad (9)$$

$$s = \sigma \cdot (\lambda T)^{-\xi} ; \qquad (10)$$

$$h = \mu + (\sigma/\xi) \cdot [(\lambda T)^{-\xi} - 1] . \qquad (11)$$

The proof can be found in [Pisarenko et al., 2009]. We see that the form parameter in the GPD and the GEV is always identical, whereas the centering and normalizing parameters differ.

Using relations (6)-(11), one can recalculate the estimates $\zeta(T), \sigma(T), \mu(T)$ obtained for some T into corresponding estimates for another time interval of different duration $\tau$:

$$\mu(\tau) = \mu(T) + (\sigma(T)/\xi) \cdot [(\tau/T)^{\xi} - 1] ; \qquad (12)$$

$$\sigma(\tau) = \sigma(T) \cdot (\tau/T)^{\xi}. \qquad (13)$$

Relations (6)-(13) are very convenient, and we shall use them in our estimation procedures. In the following, we use the notation *T* to denote the duration of a window in the known catalog (or part of the catalog) used for the estimation of the parameters, whereas we use $\tau$ to refer to a future time interval (prediction).

From the GPD-distribution (2) or the GEV-distribution (4), we can obtain the quantiles $Q_q(\tau)$, proposed as stable robust characteristics of the tail distribution of magnitudes. These quantiles are the roots of the following equations:

$$GPD_h(x | \xi, s) = q; \qquad (14)$$

$$GEV(x | \zeta, \sigma, \mu) = q. \qquad (15)$$

Inverting (14)-(15) for *x* as a function of *q*, we get:

$$Q_q(\tau) = \mu(T) + (\sigma(T)/\xi) \cdot [a \cdot (\tau/T)^{\xi} - 1] ; \quad \text{from (14)} \qquad (16)$$

$$Q_q(\tau) = h + (s/\xi)\cdot[a\cdot(\lambda\tau)^\xi - 1] \;, \qquad \text{from (15)} \qquad (17)$$

where $a = [log(1/q)]^{-\xi}$ .

## 3. Application of the GPD and GEV to the estimation of T-maximum magnitudes in Japan

### 3.1 Characteristics of the Japanese Meteorological Agency (JMA) data

The full JMA catalog covers the spatial domain delimited by $25.02 \leq latitude \leq 49.53$ *degree* and $121.01 \leq longitude \leq 156.36$ *degree* and the temporal window *01.01.1923 – 30.04.2007*. The depths of earthquakes fall in the interval $0 \leq depth \leq 657$ *km*. The magnitudes are expressed in 0.1-bins and varies in the interval $4.1 \leq magnitude \leq 8.2$. There are 39316 events in this space-time domain. The spatial domain covered by the JMA catalog covers the Kuril Islands and the East border of Asia.

Here, we focus our study to earthquakes occurring within the central Japanese islands. We thus restrict the territory of our study to earthquakes occurring within the polygon with coordinates [*(160.00; 45.00); (150.00; 50.00); (140.00; 50.00); (130.00; 45.00); (120.00; 35.00): (120.00; 30.00); (130.00; 25.00); (150.00; 25.00); (160.00; 45.00)*]. Fig.1 shows the map of the region delineated by the polygon, in which we perform our study. There were 32324 events within this area. The corresponding magnitude-frequency is shown in Fig. 2 and the histogram of magnitudes is shown in Fig. 3.

Next, we only keep "shallow" earthquakes whose depths are smaller than 70 km. We then applied the declustering Knopoff-Kagan space-time window algorithm [Knopoff and Kagan, 1977]. The remaining events constitute our "main shocks", on which we are going to apply the GDP and the GEV methods described above. There are 6497 main shocks in the polygon shown in Fig 1 with depths less than 70 km. The magnitude-frequency curve of these main shocks is shown in Fig. 4. It should be noted that the *b*-slope of the magnitude-frequency of main shocks is significantly smaller (by 0.15 or so) than the corresponding b-slope of the magnitude-frequency for all events. From the relatively small number of remaining main shocks, one concludes that the percentage of aftershocks in Japan is very high (about 80% according to the Knopoff-Kagan algorithm). The histogram of these main events with magnitudes $m \geq 5.5$ is shown in Fig. 5. One can observe irregularities and a non-monotonic behavior of histogram of magnitudes. These irregularities force us to aggregate 0.1-bins into 0.2-bins. This discreteness in the magnitudes requires a special treatment (in particular the use of Chi-square test), which is explained in the next subsection. On a positive note, no visible pattern associated with half-integer magnitude values can be detected. Thus, the use of 0.2-bins will be sufficient to remove the irregularities.

Fig. 6 plots the yearly number of earthquakes averaged over 10 years for three magnitude thresholds: $m \geq 4.1$ (all available events); $m \geq 5.5$; $m \geq 6.0$. The latter time series with $m \geq 6.0$ appears approximately stationary, with an intensity of about 3-4 events per year. *Fig .7* shows the flow of main events (same variable as Fig. 6 but for the main shocks obtained after applying the declustering Knopoff-Kagan algorithm). For large events ($m \geq 6.0$), the flow is approximately stationary.

*3.2 Adaptation for binned magnitudes*

As shown in Figs. 3 and 5, the earthquake magnitudes of the JMA catalog are discrete. Moreover, the oscillations decorating the decay with magnitudes shown in Fig. 5 require further coarse-graining with bins of 0.2 units of magnitudes as explained in the previous subsection. But, all considerations exposed in section 2 refer to continuous random values, with continuous distribution functions. For discrete random variables, the theory of extreme values is not directly applicable. This contradiction is avoided as follows.

Consider a catalog in which magnitudes are reported with a magnitude step *Δm*. Usually, in the most part of existing catalogs, including the catalog of Japan, *Δm = 0.1*. In some catalogs, two decimal digits are reported, but the last digit is fictitious unless the magnitudes are recalculated from seismic moments, themselves determined with several exact digits (such as for the $m_W$ magnitude in the Harvard catalog). Here, we assume that the digitization is fulfilled exactly without random errors in intervals *( (k-1)· Δm;  k·Δm),* where *k* is an integer. As a consequence, in the GPD approach, we should use only half-integer thresholds *h = (k-1/2)· Δm*, which is not a serious restriction.

Furthermore, having a sample of observations exceeding some *h = (k-1/2)· Δm*, and fitting the GPD-distribution to it, we need to test the goodness of fit of the GEV model to the data. For continuous random variables, the Kolmogorov test or the Anderson-Darling test was previously used successfully [Pisarenko et al., 2008; 2009]. For discrete variables, such statistical tools tailored for continuous random variables are incorrect. We calculated the Kolmogorov distances for discrete artificial sample obeying the GEV-law and found that their distribution is very far from the true one (the Kolmogorov distances for discrete magnitudes are much larger than for continuous random variables.). We are thus forced to use statistical tools adapted to discrete random variables. We have chosen the standard $\chi^2$-method, that provides both a way to estimate unknown parameters and to strictly evaluate the goodness of fit. The Chi-square test has two peculiarities:

1. In order to be able to apply the Chi-square test, a sufficient number of observations is needed in each bin (we choose this minimum number as being equal to 8 (see discussion of this matter in (Borovkov 1987);
2. In order to compare two different fits (corresponding to two different vectors of parameters), it is highly desirable to have the same binning in both experiments. Otherwise, the significance levels, which depend on the binning, can vary considerably.

In general, the Chi-square test is less sensitive and less efficient than the Kolmogorov test or the Anderson-Darling test. This results from the fact that the Chi-square coarsens data by putting them into discrete bins.

When using the GEV-approach, the digitized GEV-distribution of the magnitude maxima in successive *T*-intervals is fitted using the $\chi^2$-method.

### *3.3. The GPD approach*

Consider the discrete set of magnitudes registered with step *Δm* over threshold *h*,

$$h + (k-1)\,\Delta m/2 \leq m < h + k\Delta m/2; \quad k=1,...r\,; \qquad \Delta m = 0.1. \qquad (18)$$

The corresponding discrete probabilities read

$$p_k(\xi,s\,|h) = P\{h + (k-1)\cdot 0.05 \leq m < h + k\cdot 0.05\} =$$
$$= GPD(h + k\cdot 0.05\,|\xi,s,h) - GPD(h + (k-1)\cdot 0.05\,|\xi,s,h); \qquad (19)$$

$$p_{r+1}(\xi,s\,|h) = 1 - GPD(h + r\cdot 0.05\,|\xi,s,h). \qquad (20)$$

The last *(r+1)-th* bin covers the interval *(h + r·0.05; ∞)*. We use the following expression

$$GPD(x\,|\xi,s,h) = 1 - [1+(\xi/s)(x-h)]^{-1/\xi}, \quad h \leq x \leq h - s/\xi,\ \xi < 0. \qquad (21)$$

Let us assume that the interval (18) contains $n_k$ observations. Summing over the *r+1* intervals, the total number of observations is $n = n_1 + n_2 +... n_r + n_{r+1}$. Then, the Chi-square sum *S(ξ,s)* is written as follows:

$$S(\xi,s) = \sum_{k=1}^{r+1} [n_k - n\cdot p_k(\xi,s\,|h)]^2 / n\cdot p_k(\xi,s\,|h), \qquad (22)$$

$S(\xi,s)$ should be minimized over the parameters $(\xi,s)$. This minimum value is distributed according to the $\chi^2$-distribution with $(r-2)$ degrees of freedom. The quality of the fit of the empirical distribution by expressions (19) and (20) is quantified by the probability $P_{exc} = P\{\chi^2(r-2) \geq min(S)\}$, where $\chi^2(r-2)$ is the Chi-square random value with $(r-2)$ degrees of freedom, i.e. $P_{exc}$ is the probability of exceeding the minimum fitted Chi-square sum. The larger $P_{exc}$ is, the better is the goodness of fit.

For magnitude thresholds $h \leq 5.95$ and $h \geq 6.65$, the Chi-square sums $min(S)$ happened to be very large, leading to very small $P_{exc}$ values, indicating that such thresholds are not acceptable. For thresholds in the interval $(6.05 \leq h \leq 6.55)$, the results of the Chi-square fitting procedure are collected in Table 1. In order to obtain confidence intervals, we also performed $N_b = 100$ bootstrapping procedures on our initial sample and averaged over the obtained estimates, as described in [Pisarenko, 2008, 2009].

Table 1. Chi-square fitting procedure using the GPD approach. The parameters are estimated by minimizing $S(\xi,s)$, defined by expression (21). $M_{max}$ is the rightmost point of the magnitude distribution given by expression (3). $Q_{0.90}(10)$ is the 90% quantile of the maximum magnitude distribution ($T$-maximum magnitude) in 10-year intervals.

| $h$ | 6.05 | 6.15 | 6.25 | 6.35 | 6.45 |
|---|---|---|---|---|---|
| $r$ | 7 | 7 | 6 | 6 | 6 |
| degrees of freedom | 5 | 5 | 4 | 4 | 4 |
| $\xi$ | -0.0468 | -0.2052 | -0.2137 | -0.2264 | -0.1616 |
| $s$ | 0.5503 | 0.6420 | 0.6397 | 0.6264 | 0.6081 |
| $M_{max}$ | 17.87 | 9.43 | 9.31 | 9.11 | 10.20 |
| $Q_{0.90}(10)$ | 8.73 | 8.32 | 8.29 | 8.24 | 8.52 |
| $P_{exc}$ | 0.0753 | 0.2791 | 0.3447 | 0.3378 | 0.1747 |

As we pointed out above, if the distribution of magnitudes over thresholds obeys the $GPD(x|\xi,s,h)$, then, for a Poissonian flow of events, the $T$-maxima have the GEV-distribution:

$$GEV(x|\xi,\sigma_T,\mu_T) = exp\{-[1+(\xi/\sigma_T)(x-\mu_T)]^{-1/\xi}\}, \quad x \leq h - \sigma_T/\xi, \quad \xi < 0. \tag{23}$$

Thus, we can use an alternative approach, namely the GEV-approach, to fit the GEV-distribution to the sample of $T$-maxima derived from the same underlying catalog.

Having estimated the first triple $(\xi, \sigma_T, \mu_T)$ or the second triple $(\xi, s, h)$, we use these estimates to predict the quantile of $\tau$-maxima for any arbitrary future time interval $(0, \tau)$, since these $\tau$-maxima have the distribution $GEV(x | \xi, \sigma_\tau, \mu_\tau)$, as seen from equations (6)-(13). Recall that, in equations (6)-(13), $\lambda$ denotes the intensity of the Poissonian flow of events whose magnitudes exceed the threshold $h$.

In Table 1, three thresholds $h=6.15$: $h=6.25$ and $h=6.35$ give very close estimates. In contrast, the estimates obtained for the thresholds $h=6.05$ and $h=6.45$ have smaller goodness of fit (smaller $P_{exc}$). This suggests accepting the estimates corresponding to the highest goodness of fit $(h=6.25)$:

$$\xi_{GPD} = -0.2137; \quad s_{GPD} = 0.6397; \quad M_{max,GPD} = 9.31; \quad Q_{0.90,GPD}(10) = 8.29. \qquad (24)$$

These estimates are very close to their mean values obtained over the three thresholds $h=6.15$; 6.25; 6.35.

In order to estimate the statistical scatter of these estimates, we simulated our whole procedure of estimation $N_b = 100$ times on artificial GPD-samples with known parameters. For better stability, instead of sample standard deviations, we used the corresponding order statistics, namely, the difference of quantiles:
$$(Q_{0.84} - Q_{0.16})/2. \qquad (25)$$
For Gaussian distributions, this quantity (25) coincides with its standard deviation (std). For distributions with heavy tails, the difference (25) is a more robust estimate of the scatter than the usual std. Combining the scatter estimates (25) derived from simulations to the mean values (24), the final results of the GPD approach for the JMA catalog can be summarized by

$$\xi_{GPD} = -0.2137 \pm 0.1031; \qquad s_{GPD} = 0.6397; \pm 0.0634;$$
$$M_{max,GPD} = 9.31 \pm 1.14; \qquad Q_{0.90,GPD}(10) = 8.29 \pm 0.49; \qquad (26)$$

One can observe that the statistical scatter of $M_{max}$ exceeds the scatter of the quantile $Q_{0.90}(10)$ by a factor larger than two, confirming once more our earlier conclusion on the instability of $M_{max}$.

*3.4 The GEV approach*

In this approach, we divide the total time interval $T_c$ from 1923 to 2007 covered by the catalog into a sequence of non-overlapping and touching intervals of length $T$. The maximum magnitude $M_{T,j}$ on each $T$-interval is identified. We have $k = [T_c/T]$ $T$-intervals, so the sample of our $T$-maxima has size $k$: $M_{T,1}, ..., M_{T,k}$. We assume that $T$ is large enough, so that each $M_{T,j}$ can be considered as being sampled from the GEV-distribution $GEV(x\,|\,\xi,\sigma_T,\mu_T)$ with some unknown parameters $(\xi,\sigma_T,\mu_T)$, that should be estimated through the sample $M_{T,1},\ldots,M_{T,k}$.

The larger $T$ is, the more accurate is the GEV-approximation for this observed sample, but one cannot choose too large $T$, because the sample size $k$ of the set of $T$-maxima would be too small. This would make inefficient the statistical estimation of the three unknown parameters $(\xi,\sigma_T,\mu_T)$. Besides, we should keep in mind the restrictions mentioned above, imposed by the Chi-square method, that the number of bins should be constant for all used $T$ values and that the minimum number of observations per bin should not be less than 8. In order to satisfy these contradictory constraints, as a compromise, we had to restrict the $T$-values to be sampled in the rather small interval

$$200 \leq T \leq 300 \text{ days.} \qquad (27)$$

It should be noted that, for all $T$-values over $50$ days, the estimates of the parameters do not vary much and, only for $T \leq 40$, do the estimates change drastically. We have chosen $T=200$ and obtained the following estimates:

$$\xi_{GEV} = -0.1901 \pm 0.0717; \quad \mu_{GEV}(200) = 6.3387 \pm 0.0380; \quad \sigma_{GEV}(200) = 0.5995 \pm 0.0223;$$
$$M_{max,GEV} = 9.57 \pm 0.86; \qquad Q_{0.90,GEV}(10) = 8.34 \pm 0.32; \qquad (28)$$

The estimates of the scatter in (28) were obtained by the simulation method with 100 realizations, similarly to the method used in the GPD approach. In estimating the parameters, we have used the *shuffling* procedure described in [Pisarenko et al., 2009], which is similar to the bootstrap method, with $N_S = 100$ realizations. It should be noted that, in equation (28), the $T$-value for the parameters $\mu,\sigma$ is indicated in days $(T=200$ days$)$ whereas, in the quantile $Q$, the $\tau$-value is indicated in years ($\tau = 10$ years).

Comparing $\xi$, $M_{max}$ and the Q-estimates obtained by the GPD and the GEV approaches, the GEV-method is found to be somewhat more efficient (its scatter is smaller by a factor approximately equal to 0.7). This can be explained by the fact that the GEV-approach uses the full catalog more intensively: all events with magnitude $m \geq 4.1$ participate (in principle) in the estimation, whereas the GPD-approach throws out all events with $m < h$.

Finally, we show in Figs. 8 and 9 the dependence of the quantile $Q_q(\tau)$ as a function of $\tau$, for $\tau = 1\div 50$ years, as estimated by our two approaches, respectively given by expressions (16) and (17). One can observe that the quantile $Q_q(\tau)$ obtained by the two methods are very close. This testifies on the stability of the estimations. Fig. 10 plots the median (quantile $Q_q(\tau)$ for $q =$

*50%*) of the distribution of the maximum magnitude as a function of the future $\tau$ years, together with the two accompanying quantiles 16% and 84% corresponding to the usual ± one standard deviations. These quantiles $Q_q(\tau)$ can be very useful tools for pricing risks in the insurance business and for optimizing the allocation of resources and preparedness by state governments.

*4. Discussion and conclusions.*

We have adapted the new method of statistical estimation suggested in (Pisarenko et al., 2009) to earthquake catalogs with discrete magnitudes. This method is based on the duality of the two main limit theorems of Extreme Value Theory (EVT). One theorem leads to the GPD (peak over threshold approach), the theorem leads to the GEV (*T*-maximum method). Both limit distributions must possess the same form parameter $\xi$. For the Japanese catalog of earthquake magnitudes over the period 1923-2007, both approaches provide almost the same statistical estimate for the form parameter, which is found negative; $\xi \cong -0.2$. A negative form parameter corresponds to a distribution of magnitudes which is bounded from above (by a parameter, named $M_{max}$). This maximum magnitude corresponds to the finiteness of the geological structures supporting earthquakes. The density distribution extends to its final value $M_{max}$ with a very small probability weight in its neighborhood, characterized by a tangency of a high degree ("duck beak" shape). In fact, the limit behavior of the density distribution of Japanese earthquake magnitudes is described by the function *$(M_{max}-m)^{-1-1/\xi} \cong (M_{max}-m)^4$*, i.e. by a polynomial of degree approximately equal to 4. This is the explanation of the unstable character of the statistical estimates of the parameter $M_{max}$: a small change of the catalog of earthquake magnitude can give rise to a significant fluctuation of the resulting estimate of $M_{max}$. In contrast, the estimation of the integral parameter $Q_\tau(q)$ is generally more stable and robust, as we demonstrate quantitatively for the Japanese catalog of earthquake magnitudes over the period 1923-2007.

The main problem in the statistical study of the tail of the distribution of earthquake magnitudes (as well as in distributions of other rarely observable extremes) is the estimation of quantiles, which go beyond the data range, i.e. quantiles of level *q > 1 – 1/n*, where n is the sample size. We would like to stress once more that the reliable estimation of quantiles of levels *q > 1 – 1/n* can be made only with some additional assumptions on the behavior of the tail. Sometimes, such assumptions can be made on the basis of physical processes underlying the phenomena under study. For this purpose, we used general mathematical limit theorems, namely, the theorems of EVT. In our case, the assumptions for the validity of EVT boil down to assuming a regular (power-like) behavior of the tail $1 - F(m)$ of the distribution of earthquake magnitudes in the vicinity of its rightmost point $M_{max}$. Some justification of such an assumption can serve the fact that, without them, there is no meaningful limit theorem in EVT. Of course, there is no a priori guarantee that these assumptions will hold in some concrete situation, and they should be discussed and possibly verified or supported by other means. In fact, because EVT suggests a statistical methodology for the extrapolation of quantiles beyond the data range, the question whether such interpolation is justified or not in a given problem should be investigated carefully in each concrete situation. But EVT provides the best statistical approach possible in such a situation.

*Acknowledgements.* This work was partially supported (V.F.Pisarenko, M.V.Rodkin) by the Russian Foundation for Basic research, grant 09-05-01039a, and by the Swiss ETH CCES project EXTREMES (DS).

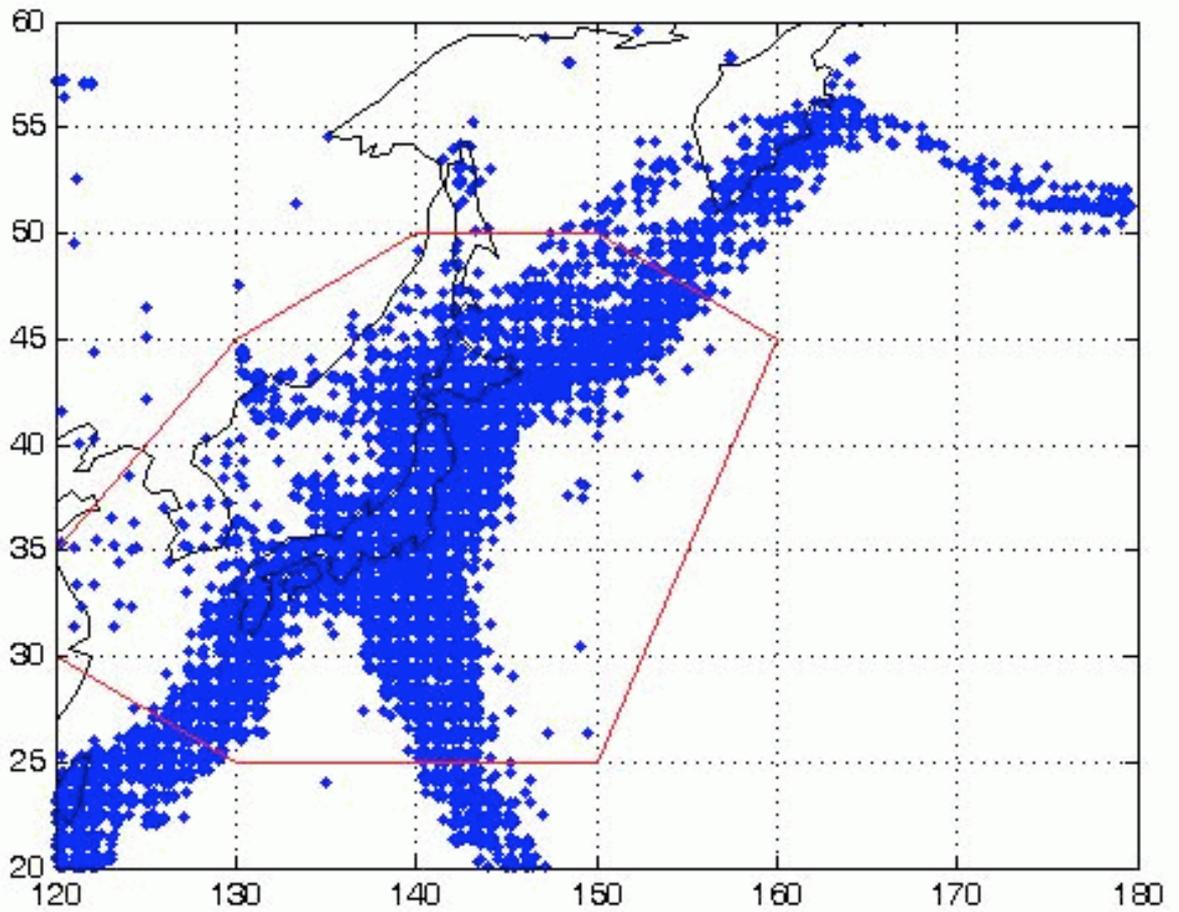

**Fig. 1:** Map of the region kept for our study; the coordinates of nodes of the polygon delimiting the area of study are [*(160.00; 45.00); (150.00; 50.00); (140.00; 50.00); (130.00; 45.00); (120.00; 35.00): (120.00; 30.00); (130.00; 25.00); (150.00; 25.00); (160.00; 45.00)*].

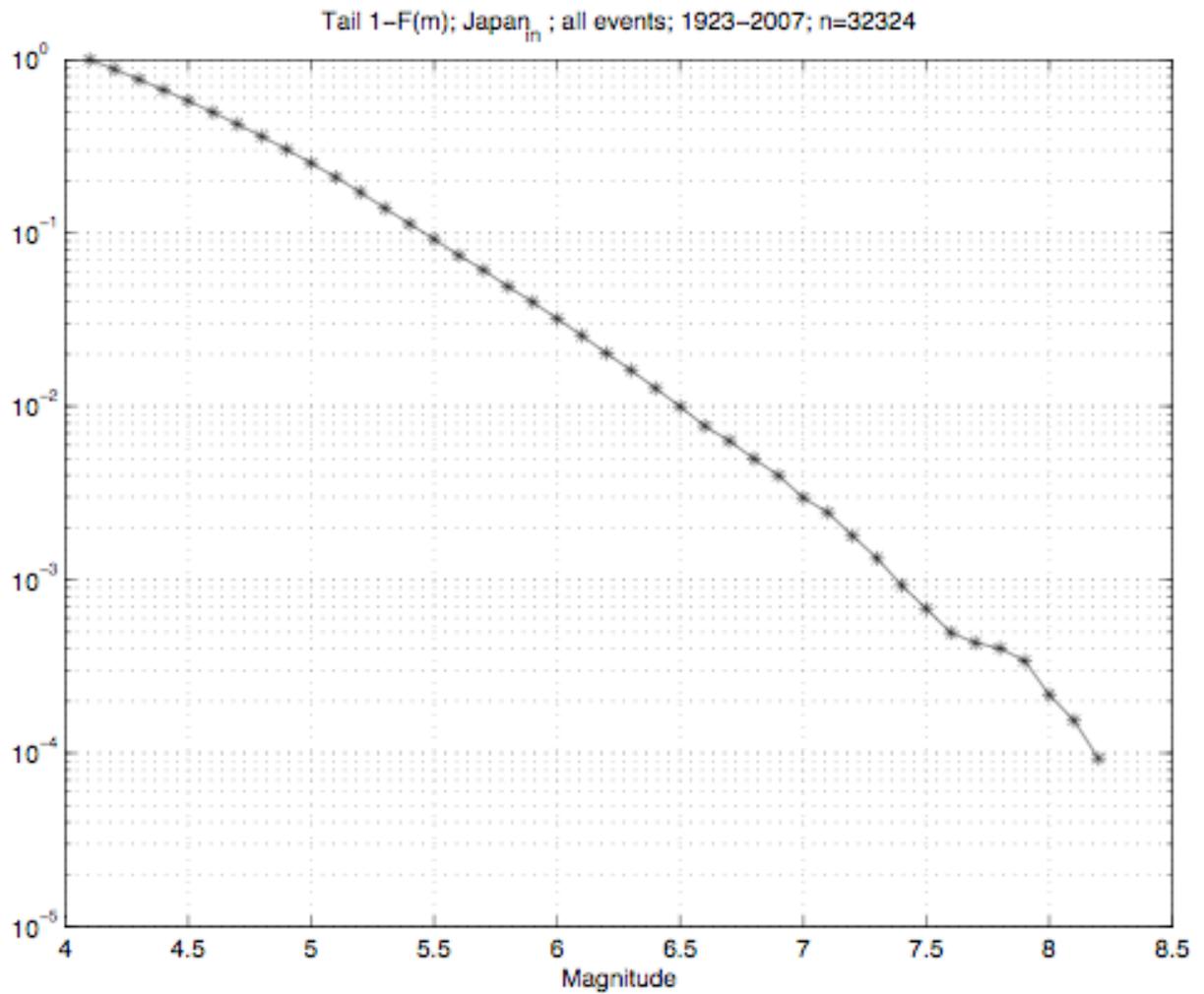

**Fig. 2:** Magnitude-frequency distribution of the 32324 earthquakes that occurred in the region delimited by the polygon shown in Fig. 1 over the period from 1923 to 2007.

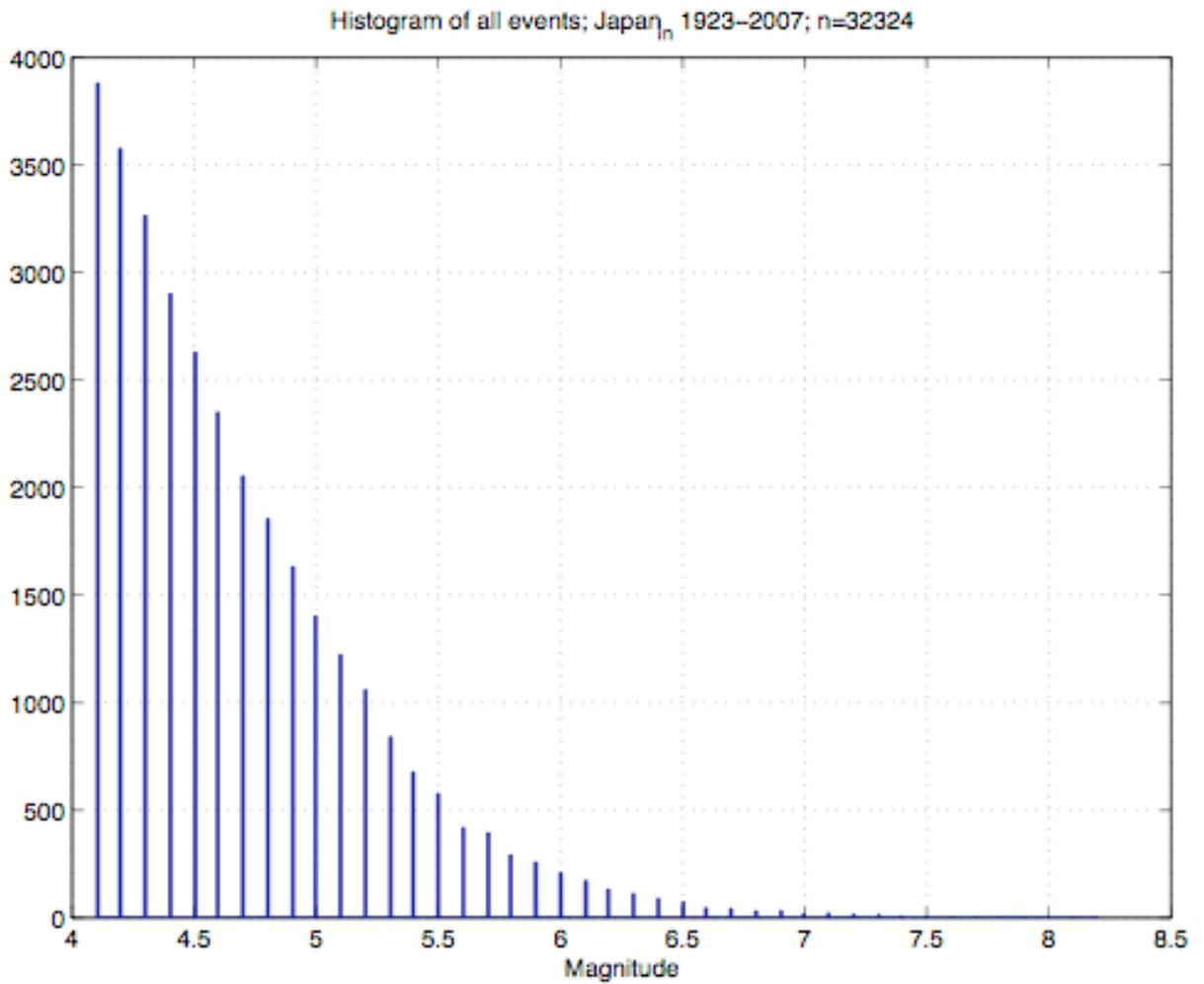

**Fig. 3:** Histogram of the magnitudes of earthquakes used in Fig. 2. The discrete 0.1 bins are clearly visible.

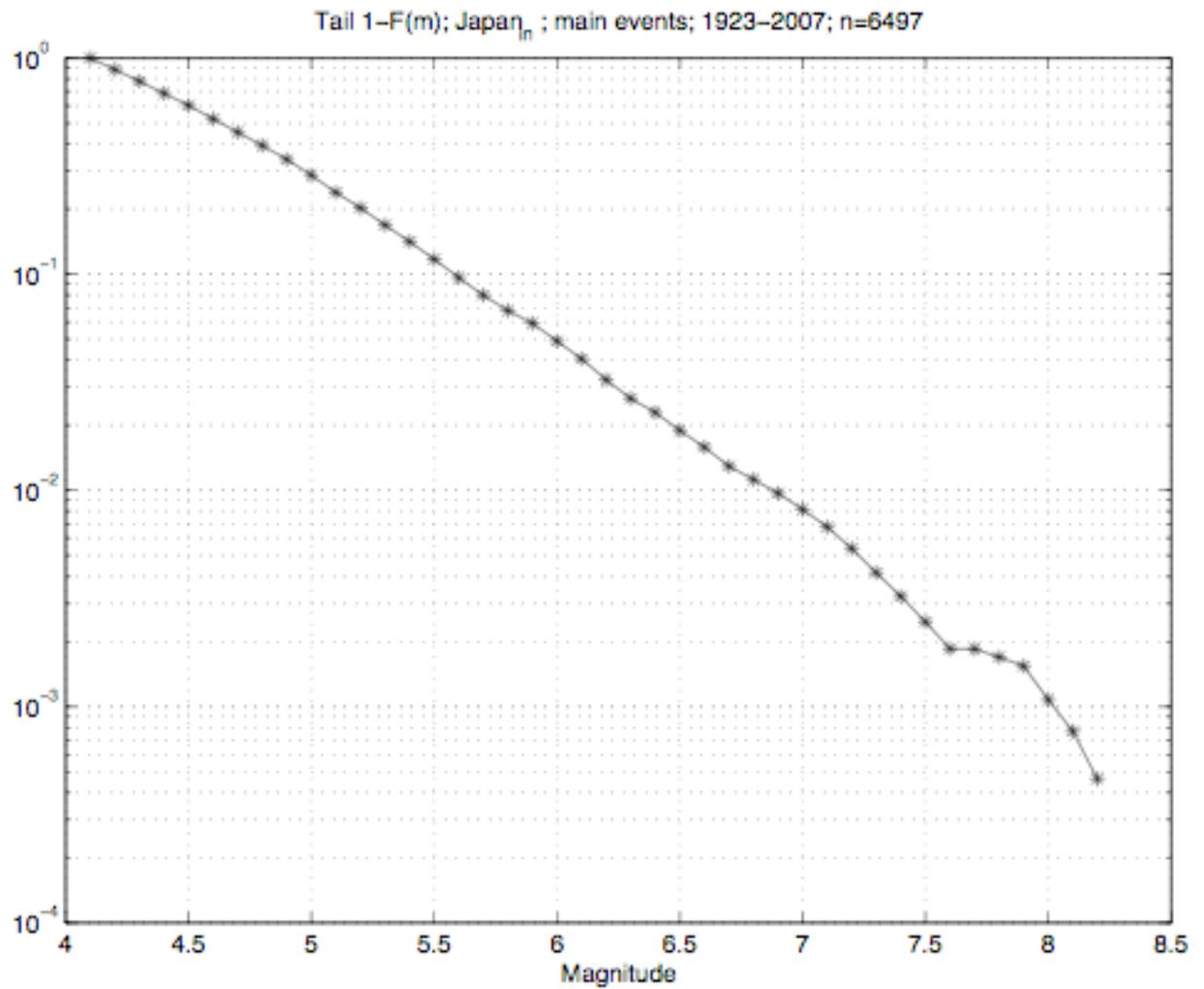

**Fig. 4:** Magnitude-frequency distribution of the 6497 "main shocks" remaining in the domain delineated by the polygon shown in Fig. 1 over the period from 1923 to 2007, which are of depths smaller than 70 km, after applying the Knopoff-Kagan declustering algorithm.

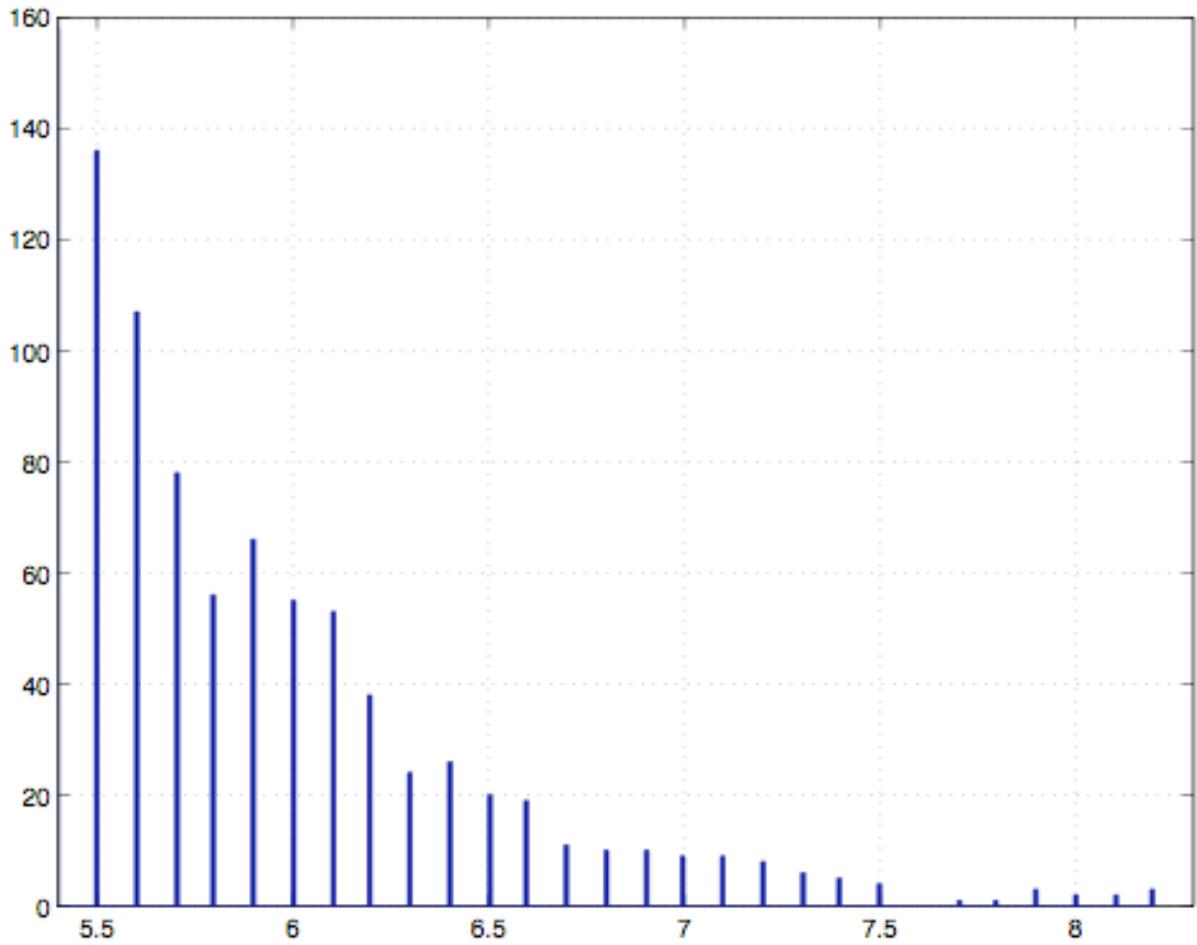

**Fig. 5:** Histogram of the magnitudes of the main shocks whose magnitude-frequency distribution is shown in Fig. 4. The discrete 0.1 bins are clearly visible. There are additional oscillations decorating the decay with magnitudes, which require further coarse-graining, as explained in the text.

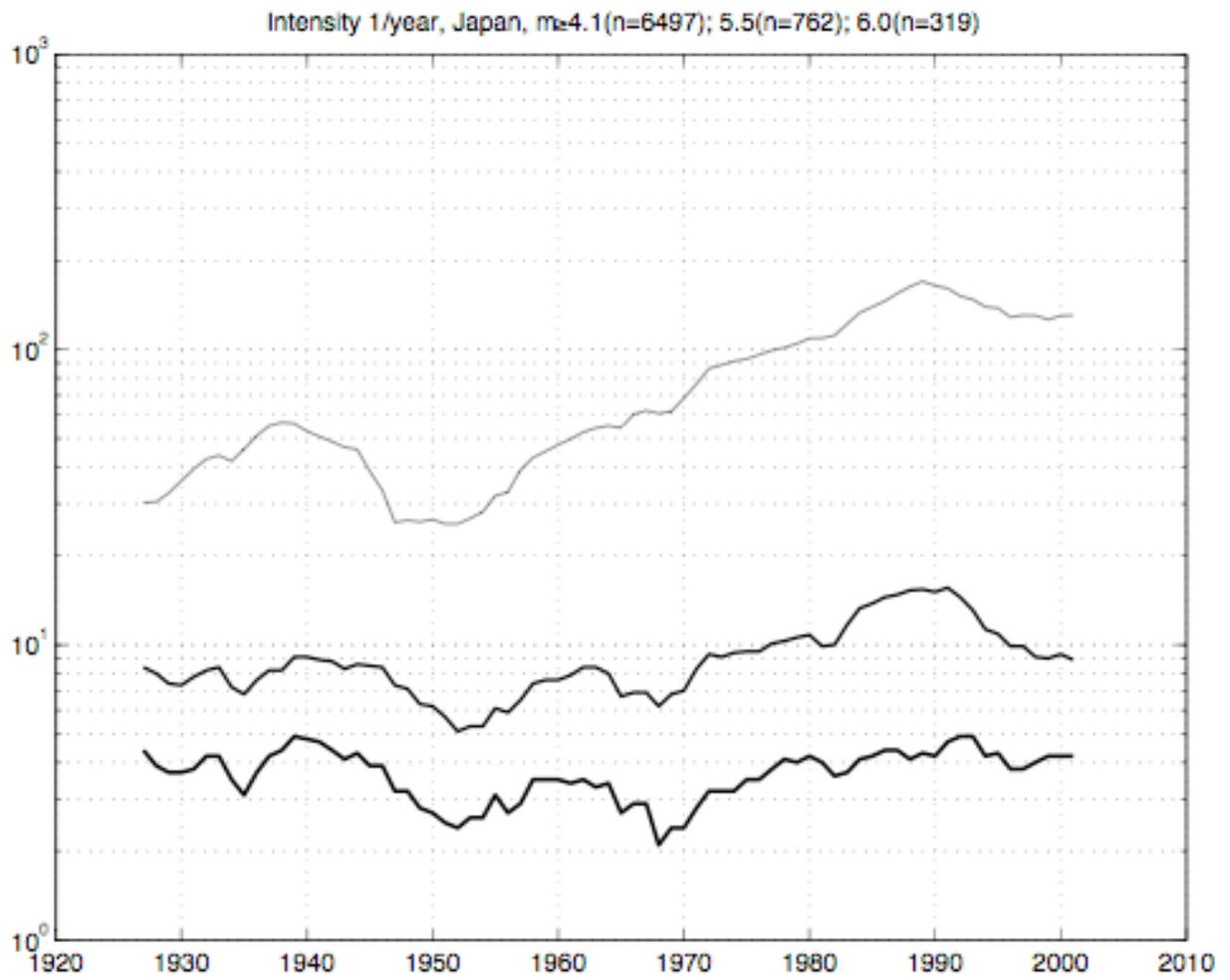

**Fig. 6:** Yearly number of earthquakes averaged over 10 years for three magnitude thresholds: $m \geq 4.1$ (all available events); $m \geq 5.5$; $m \geq 6.0$.

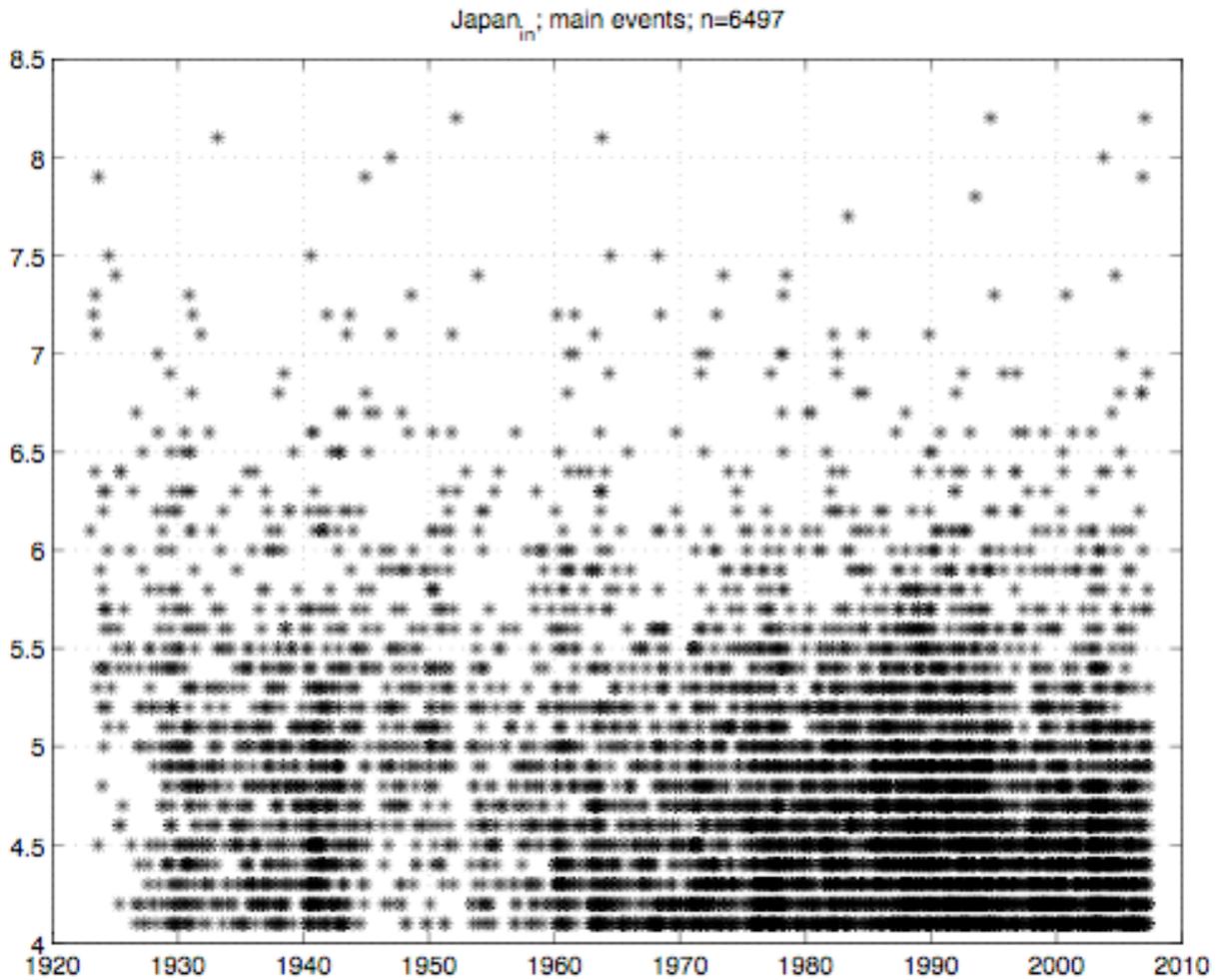

**Fig. 7:** Flow of main shocks from 1923 to 2007. Main shocks are defined as "shallow" earthquakes inside the polygonal domain shown in Fig. 1, whose depths are smaller than 70 km and which remain after applying the declustering Knopoff-Kagan space-time window algorithm [Knopoff and Kagan, 1977].

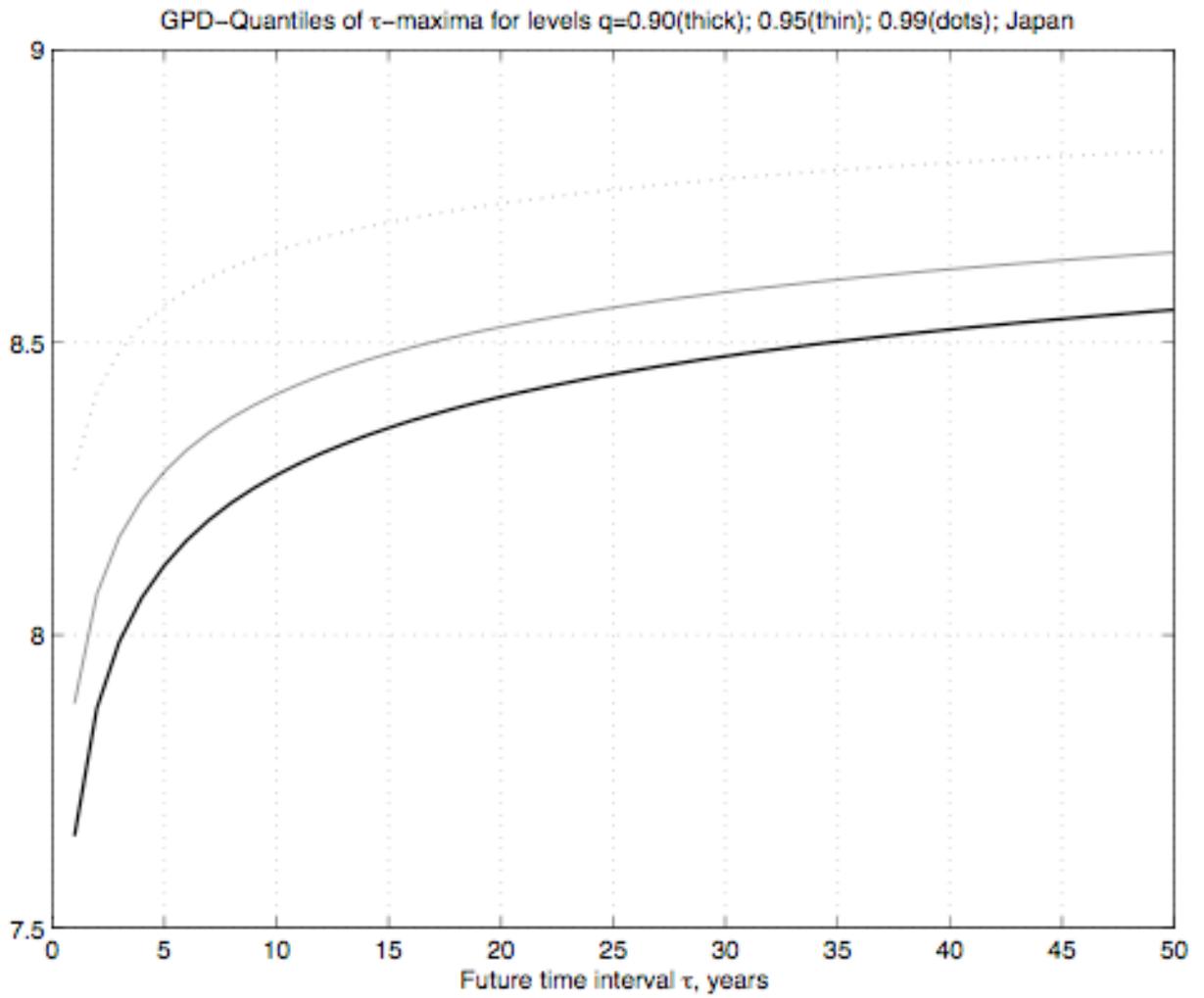

**Fig. 8:** Quantile $Q_q(\tau)$ of the distribution of maxima over a future time interval $\tau$, for three confidence levels $q$, defined by expression (16). The three curves use the parameters of the GDP estimated from the JMA catalog, as explained in the text.

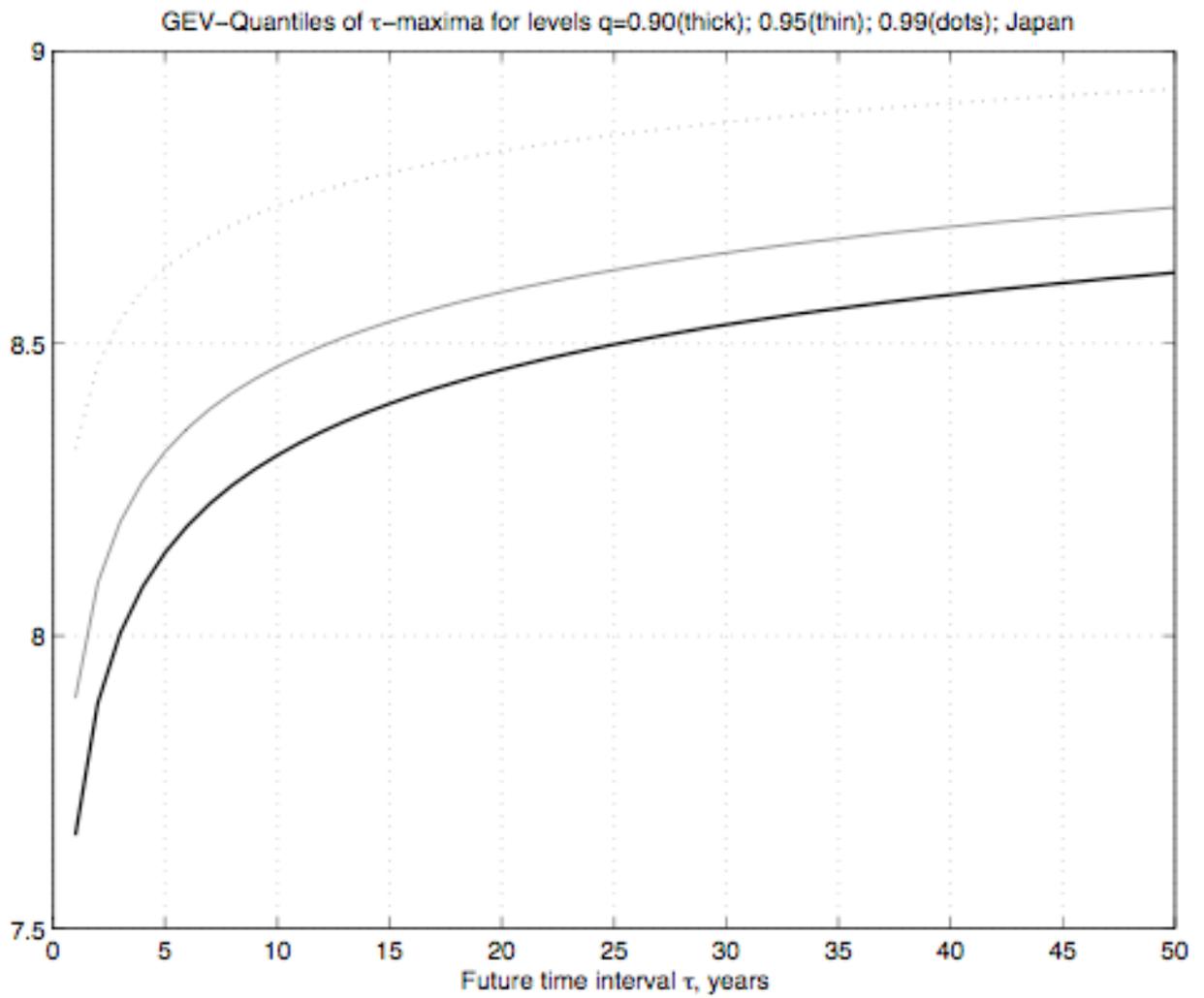

**Fig. 9:** Quantile $Q_q(\tau)$ of the distribution of maxima over a future time interval $\tau$, for three confidence levels $q$, defined by expression (17). The three curves use the parameters of the GEV estimated from the JMA catalog, as explained in the text.

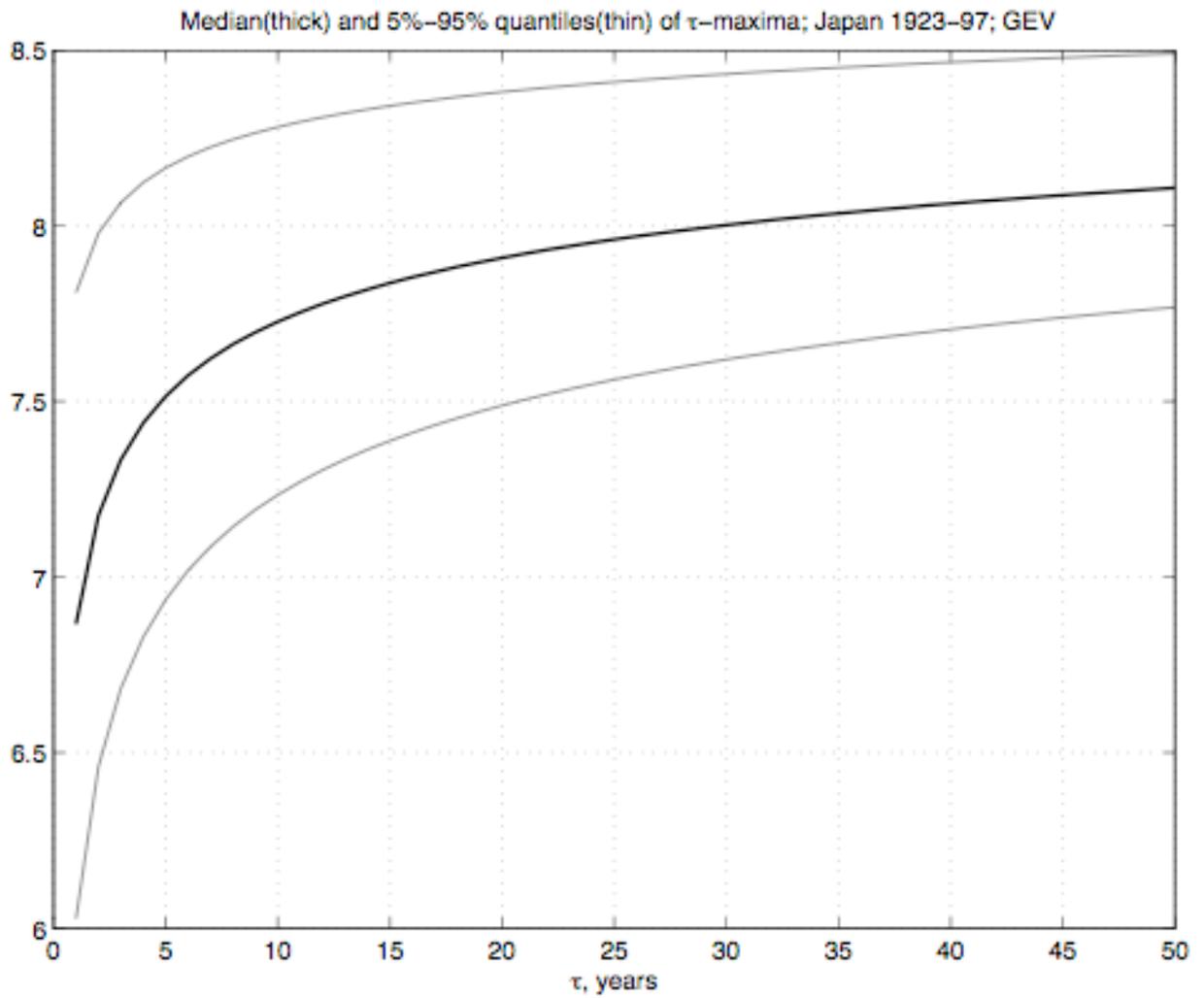

**Fig. 10:** Median (quantile $Q_q(\tau)$ for $q = 50\%$) of the distribution of the maximum magnitude over a future time interval $\tau$, obtained by the GEV method, as a function of $\tau$ (years), together with the two accompanying quantiles 16% and 84% corresponding to the usual ± one standard deviations.